\documentstyle[12pt]{article}
\begin{document}
\title{
\begin{flushright}
{\small SMI-94-2 \\
 CVR-94-10}
 \end{flushright}
Quantum Group Sheaf and Quantum Manifolds}
\date {~}
\author{ I.V.Volovich
\thanks{On leave of absence from Steklov Mathematical
institute, Vavilov St.42, 117966, Moscow}
\\ Centro Matematico V.Volterra \\
Universita di Roma Tor Vergata}
\maketitle
\bigskip\bigskip
\centerline{ {\bf Abstract } }
The problem of introducing a dependence of elements
of quantum group  on classical parameters is considered.
It is suggested to interpret a homomorphism from the
algebra of functions on quantum group to the algebra
of sections of a sheaf of algebras on a classical
manifold as describing such a dependence.
It is argued that the functorial point of view
of group schemes is more appropriate in quantum group
field theory.

A sheaf of
the Hopf algebras over the manifold (quantum sheaf)
is constructed by using  bosonization formulas
for the algebra of functions on the quantum group  $SU_{q}(2)$
and the theory of representations
of canonical commutation relations.
A family of automorphisms
of the Hopf algebra depending on classical variables
is described.

Quantum manifolds, i.e. manifolds
with commutative and non-commutative coordinates
are discussed as a generalization of
supermanifolds.

Quantum group chiral fields
and relations with algebraic
differential calculus are discussed.
\newpage
\medskip

\bigskip\bigskip\medskip
\centerline{{\bf 1. \quad Introduction}}

\bigskip
There are  numerous
applications of quantum groups. One of approaches
uses quantum group as a deformation of
gauge group in the theory of gauge fields
preserving the standard classical
space-time \cite{1}-\cite {10}.

The gauge function $g(x)$ in the standard theory of
gauge fields is a map $g:M\to G$
from space-time $M$ to a Lie group $G$. This function
$g(x)$  can also be considered as a chiral field.
There is a problem how  to give a mathematical
meaning to a function  $g:M\to G_q$
from the classical manifold $M$ to quantum group $G_q$.
This problem was discussed in   \cite{1,2,9,10,7}.

The purpose of this note is to suggest a construction
which can be interpreted as giving
a dependence of elements of quantum group on classical
parameters.

Let us recall that if one has a continuous
map $g:M\to G$ from a topological space $M$ to a topological
space $G$ then there exist a natural homomorphism
$F:A(G)\to A(M)$ of algebras of continuous
functions corresponding to the map $g$. If instead of $A(G)$
one considers the noncommutative algebra of functions
on a quantum group $A(G_q)$ then there is not a nontrivial
homomorphism from  $A(G_q)$ to the commutative algebra $ A(M)$.
Therefore we will consider a homomorphism $F$ from  $A(G_q)$
to the algebra of sections $B(M)$ of a sheaf $B$
of noncommutative algebras over $M$ and we will interpret
$F$ as giving
a dependence of elements of quantum group on classical
parameters from $M$. This definition involves the sheaf $B$
and one can ask about the most natural choice  of $B$.
We will construct in this paper some examples of such
sheaves. From the other hand if one works with an arbitrary
sheaf $B$ then one has the functoriality of the construction
and it is more sensitive to speak about dependence
of elements of quantum group on classical parameters.
In fact we are using the idea of group schemes and representative
functor [44-47,57] from algebraic geometry.

    It will be introduced a family (quantum sheaf)
of algebras $\{A_x\}$ parametrized by a point $x\in M$,
where  $A_x$, for a given $x$,  is isomorphic to the
Hopf algebra
of functions on the quantum group $SU_q(2)$. The map
$g:M\to SU_q(2)$ is interpreted as a section of this quantum sheaf.
There are
nontrivial commutation relations  between elements from
$A_x$ and from $A_y$ for different $x$ and $y$.
All algebras  $A_x$   are realized as
algebras of operators    in a Hilbert space
which is constructed by means of the theory
of representations of canonical commutation
relations. In the simplest case one has the Fock
representation.

Our construction is based on a generalization
of representation of generators of the algebra of functions
on the quantum group  $SU_q(2)$ in terms of creation
and annihilation operators from  \cite{11}. The bosonization
from   \cite{11} is an analogue of the known
bosonization formulas for generators of the q-deformation
of the universal enveloping of the Lie algebra of $SU(2)$
 \cite{12}-\cite{14}.
We will present also an infinite dimensional generalization
of these formulas which could be useful for construction
of q-deformed gauge fields.

Note here that
just simple tensor product  $\otimes_x A_x$ of the  same
algebra $A_x =A$ is not suitable for our purpose
of receiving the smooth  function $g(x)$.

We will discuss also a possible relation of our
construction with the algebraic differential calculus
on quantum groups  \cite{17}-\cite{26},\cite{1}-\cite{10}.

Quantum manifolds, i.e.manifols with commuting and
noncommuting coordinates will be discussed
as a generalization of supermanifolds.

\newpage
\centerline {\bf 2. Quantum Group Sheaf and Group Schemes}
\bigskip

Let us recall some definitions and introduce
notations. We will be interested in the notion of
a sheaf of algebras on a topological space $M$, see for example
\cite{34}.  Let $K$ will be a field of real numbers $R$ or
complex numbers $C$. We will consider algebras over the
field $K$. Recall that presheaf $B$ of algebras on $M$
is a contravariant functor from the category of open
subsets of $M$ and inclusions to the category of algebras.
This means that a presheaf $B$ on $M$ is a function
which assigns, to each open set $U\subset M$, an algebra
$B(U)$ and which assigns, to each pair $U\subset V$
of open sets, a homomorphism of restrictions
$r_{U,V} :B(U)\to B(V)$ in such a way that
$r_{U,U}=1$ and $r_{U,V} r_{V,W}=r_{U,W}$
when $U\subset V\subset W$. For example one has the presheaves
of differential functions on a differential manifold $M$;
of vector fields on $M$ and so on. In the last example
the algebra $B(U)$ of vector fields on $U$ is a noncommutative
one (it is a Lie algebra). Elements from $B(U)$ are called
sections over $U$.

A presheaf $B$ is called a sheaf if the following conditions
are satisfied:

(i) If $U=\cup U_{\alpha}$, with $U_{\alpha}$ open in $M$,
and $s\in B(U)$ such that $s|U_{\alpha}$ for all ${\alpha}$,
then $s=0$.

(ii) Let $\{U_{\alpha}\}$ be a collection of open sets
in $M$ and let $U=\cup U_{\alpha}$. If $s_{\alpha}\in
B(U_{\alpha})$ are given such that
$s_{\alpha}|U_{\alpha}\cap U_{\beta}=s_{\beta}|
U_{\alpha}\cap U_{\beta}$ for all $\alpha ,\beta$
then there exist an element $s\in B(U)$ with
$s|U_{\alpha}=s_{\alpha}$ for all $\alpha$.

The stalk of germs of $B$ at a point $x\in M$ inherits
a natural algebraic structure from the algebras $B(U), U\ni x$.
Any presheaf $B$ generates a covering space and a sheaf
of sections of the space. The algebra of sections of the sheaf
$B$ over $M$ is denoted by $B(M)$.

If $M$ and $G$ are topological spaces and $f: M\to G$ is
a continuous map then one has a homomorphism
$F=F_f:A(G)\to A(M)$ of algebras of continuous
functions defined by the formula
$F(g)=g\circ f$ where $g\in A(G)$. One interprets
such a homomorphism as describing a dependence
of elements of $G$ on variables from $M$.

If $B$ and $A$
are presheaves on $M$ and $G$ respectively, then
an $f$-cohomomorphism  $F:A\to B$ is a family of
homomorphisms $F_{U}:A_{U}\to B(f^{-1}(U))$
for $U$ open in $G$, compatible with restrictions.

If $A(G)$ is a noncommutative algebra, for example
the algebra of functions on a quantum group $A(G_q)$, then
there are not nontrivial homomorphisms from $A(G_q)$
to the commutative algebra  $A(M)$. However one can
consider a sheaf $B$ of noncommutative algebras on $M$
and interpret a homomorphism $F:A(G_q)\to B(M)$ as defining
a dependence of elements of quantum group on classical
parameters.

{\it Quantum group sheaf} is a quadruple $(A,M,B,F)$ where
$A$ is the algebra of functions on a quantum group,
$M$ is a topological space, $B$ is a sheaf of algebras
over $M$ and $F:A\to B(M)$ is a homomorphism of
the algebra $A$ to the algebra of sections of $B(M)$.
One can take also the trivial sheaf $A$ over $M$
and consider a homomorphism of sheaves $F:A\to B$.

We interpret the homomorphism $F$ as introducing
a dependence of elements of quantum group on
'classical' parameters from $M$. For example
if $a,c,...$ are generators of the algebra
of functions on the quantum group $SU_q(2)$
(see below) satisfying to relations $ac=qca,...$,
then $F(a)$ and $F(c)$ are sections of the sheaf
$B$ and one has relations $F(a)(x)F(c)(x)=qF(c)(x)F(a)(x)$,
for $x\in M$.

Another example will be given by a map from a manifold
$M$ to the group of automorphisms of the Hopf algebra $A$,
$x\to F(a)(x)=U_x aU_x^{-1}$, see below.

There are essentially two points of view in modern
algebraic geometry, see \cite{45}. Let us take a simple
example: If $P_1,...,P_m$ are polynomials in
$n$ indeterminates, we may assign to them,
on the one hand the subset $X$ of $C^n$ consisting
of the points $x$ such that $P_1 (x)=...=P_m (x)=0$
-this is the geometrical point of view. On the other hand
we may watch the functor assigning to every commutative
associative algebra $A$ the set $X(A)$ formed by all
$x\in A^n$ such that $P_1 (x )=...=P_m (x)=0$- this is
the functorial point of view. Now instead of defining schemes
as geometric spaces endowed with sheaves of local rings,
one defines schemes as functors over some category of
rings and then show that the category of these functors
is equivalent to some category of geometric spaces.

We deal with a set of homomorphisms $Hom(A,B)$
from the algebra $A$ to the algebra $B(M)$
of sections of the sheaf $B$ of algebras over $M$.
For a given $A$ and $M$ one has a functor
$B\to Hom(A,B)$ from the category of sheaves of algebras
over $M$ to the category of sets. This functor is an
extension of the following known construction.
Let $G$ be an affine algebraic group, $A=K[G]$
is the ring of regular functions on $G$ and $B$
is an arbitrary $K$-algebra. Then the functor
$B\to G(B)=Hom(A,B)$ is in fact
a functor from the category of $K$-algebras
to the category of groups. It was introduced
by A.Grothendieck and it is called
the group scheme over $K$, see \cite{44}-\cite{46},\cite{57}.
An approach to quantum groups in terms of group schemes
has been actually used in \cite{1} for some
special cases. A general case was mentioned  in \cite{48}.

\bigskip
\centerline  {\bf 3. Bosonization of  $SU_q(2)$}
\bigskip
Let us recall that the algebra of functions $A=Fun(SU_q(2))$
on the quantum group $SU_{q}(2)$,  $0<q<1$
is the Hopf algebra which is generated by the elements
$a,c,a^{*}$ and $c^{*}$ and relations  \cite{27,28,18,49}:
$$
ac=qca,~~ ca^{*}=qa^{*}c,~~c^{*}c=cc^{*},, \eqno(1)
$$

$$
aa^{*}+q^{2}c^{*}c=1,~~ a^{*}a+cc^{*}=1.
$$ 

Consider a matrix $g$ of the following  canonical form
$$
g=\left( \begin{array}{rr}
a & -qc^{*}\\
c & a^{*}
\end{array} \right)
$$
 Then the
requirement of the  unitarity condition
$$
gg^{*}=g^{*}g=I
$$
is equivalent to the relations  (1).
The coproduct $\Delta :A\to A\otimes A$ is
defined by the standard formula:
$$ 
\Delta (g_{ij})=\sum g_{ik} \otimes g_{kj}
$$    

The relations  (1)
have a
representation  \cite{11} in terms of standard creation and
annihilation operators $\alpha$ and $\alpha^{*}$ satisfying
to the canonical commutation relations

$$
[\alpha,\alpha^{*}]=1  \eqno (2)
$$
If one puts
$$
a=(1-q^{2(N+1)})^{1/2 }(N+1)^{-1/2}\alpha  \eqno (3)
$$
$$
c=e^{i\theta}q^{N}
                                            \eqno (4)
$$
where  $\theta$ is a real number and
$$
N=\alpha^{*}\alpha
                                           \eqno (5)
$$
then relations (1) are satisfied.
If one defines $q=e^{-\beta}$, $\beta >0$, then
$$ 
q^N=e^{-\beta N}
$$    

One can define the operators $\alpha$ and $\alpha^{*}$
in terms of $a$ and $c$ as follows:
$$
N=-\frac{1}{2\beta}\log c^{*}c,
$$

$$ 
\alpha = (1-q^{2(N+1)})^{1/2 }(N+1)^{-1/2}a
$$    

Therefore if one ignores the known problems with unboundedness
of creation and annihilation operators one has the
following

{\bf Proposition 1}. {\it There is a one to one correspondence
(modulo dilatation} $c\to e^{i\theta}c$) {\it between
representations of the algebra }$A$    {\it
of functions on the quantum group }$SU_{q}(2)${\it
and representations of canonical commutation
relations  } (2).

One can use this proposition to reduce the theorem
on classification of irreducible representations
of $A$   \cite{29,30} to the known Stone - von Neumann theorem
on uniqueness of irreducible representations
of canonical commutation relations in the Weyl form  \cite{31}-\cite{33}.

Now we construct a family of algebras $A_f$
with generators $a_f,~c_f~,a_f^{*},~c_f^{*}$
parametrized by elements $f$ from a Hilbert space.

\bigskip
\centerline  {\bf 4. Representations of canonical commutation relations
and quantum groups}
\bigskip
Let $H$ be a complex pre-Hilbert space with inner
product $(\cdot ,\cdot )$. A representation of the
canonical commutation relations (CCR) over $H$
is a map $f\to W(f)$ of $H$ into the unitary
operators $U({\cal H})$ on a Hilbert space ${\cal H}$
satisfying

$$ 
W(f_1)W(f_2)=exp \{ \frac{i}{2} Im (f_1,f_2) \} W(f_1 + f_2)
$$
such that for each $f\in H$ the map $\lambda \to W(\lambda f)$
of $R$ into $U({\cal H})$  is strongly continuous   \cite{31,33}.
There exist self-adjoint operators $R(f)$ such that
$W(f)=\exp \{iR(f)\}$. One constructs the annihilation and
creation operators:
$$ 
b(f)=\frac{1}{\sqrt 2}(R(f)+iR(if)),
$$    

$$ 
b^{*}(f)=\frac{1}{\sqrt 2}(R(f)-iR(if))
$$    
They satisfy

$$ 
[~b(f_1),b^{*}(f_2)~]=(f_1,f_2)
$$    

For a cyclic representation with a vacuum vector $\Omega$
one introduces the generating functional $\mu (f)$
defined by
$$ 
\mu (f)=(\Omega ,W(f)\Omega)
$$    
The generating functional characterizes the
representation of CCR. In particular
$$ 
\mu (f)=exp(-\frac{1}{4}||f||)
$$    
corresponds to the Fock representation.

Now for any non-zero $f$ from $H$ let us define
$$ 
\alpha_f=\frac{b(f)}{||f||},~\alpha_f^{*}=\frac{b(f)^{*}}{||f||}
$$    
Then one has
$$ 
[~\alpha_f,\alpha_f^{*}~]=1
$$    
and one can use the bosonization formulas (3)-(5)
to construct operators
$$ 
a_f=(1-q^{2(N_f +1)})^{1/2 }(N_f +1)^{-1/2}\alpha_f ,
$$    

$$ 
c_f=e^{i\theta_f}q^{N_f},~~N_f=\alpha_f^{*} \alpha_f
$$    
satisfying the relations (1)
$$ 
a_f c_f=qc_f a_f
$$    
etc.  These operators generate the Hopf algebra $A_f$.
One has the family of unitary matrices $g_f$,
$$ 
g_f=\left( \begin{array}{rr}
a_f & -qc^{*}_f\\
c_f & a^{*}_f
\end{array} \right),
$$    

$$ 
g_f g^{*}_f =g^{*}_f g_f =I
$$    
For $f=0$ one puts $\alpha_f =\alpha$.
We have proved the following

{\bf Proposition 2.} {\it There exists a family } ${\cal A}=\{ A_f\}$
{\it of the Hopf
algebras of operators in the Hilbert space} ${\cal H}$ {\it such
that for any}  $f$ {\it from the pre-Hilbert space} $H$
{\it the algebra }$A_f$ {\it is isomorphic to the algebra }$A$ {\it of
functions on the quantum group } $SU_{q}(2)$.

It would be interesting to give an
axiomatic characterization of the quantum sheaf  $\cal {A}$.
To this end one needs to describe algebraic relations
between elements from $A_f$ and $A_h$ that could give
properties of the transition functions of the bundle.

\bigskip
\centerline  {\bf 5. Localization}
\bigskip

Up to now the pre-Hilbert space $H$ was an abstract space.
To relate our construction with the classical manifold $M$
one can consider an arbitrary map from $M$ to $ H$, i.e.
one considers a function $f_x$ depending on $x\in M$ and taking
values in $H$. Then operators
$$ 
\alpha (x)=\frac{b(f_x)}{||f_x ||},~\alpha (x)^{*}=\frac{b(f_x )^{*}}
{||f_x ||}
$$    
satisfy the relations

$$ 
[\alpha(x),\alpha(x)^{*}]=1                        \eqno  (6)
$$    
and we can use them to construct generators $a(x),~c(x)$ of a
Hopf algebra depending on the variable $x\in M$.
To be more concrete let us take $M=R^d, H=L^2 (R^d )$, fix a non-zero
function $f\in H$ and define
$$ 
\alpha (x)=\frac{b(f(x-\cdot )}{||f||},~\alpha (x)^{*}=
\frac{b(f(x-\cdot )^{*}}{||f||}
$$    
If one uses the standard notations such as
$$ 
b(f)=\int b(u)f(u)du
$$    
$$[b(u),b^{*}(u')]=\delta (u-u')$$
then
$$ 
\alpha(x)=\frac{1}{||f||} \int b(u)f(x-u)du
$$    
We get commutation relations (6) and one defines operators
$$ 
a(x)=(1-q^{2(N(x) +1)})^{1/2 }(N(x) +1)^{-1/2}\alpha(x),
$$    
$$ 
c(x)=e^{i\theta (x)}q^{N(x)},~~N(x)=\alpha(x)^{*} \alpha(x)
$$    
satisfying relations (1) depending on $x$:
$$ 
a(x) c(x)=qc(x) a(x)
$$    
etc. Here $\theta (x)$ is a real valued
function.

Therefore we have  constructed a map  $g:R^d\to SU_{q}(2)$,
$$ 
g(x)=\left( \begin{array}{rr}
a(x) & -qc^{*}(x)\\
c(x) & a^{*}(x)
\end{array} \right)
$$    

and one has the family of algebras  $\cal {A}=\{A_x\}$ for $x\in R^d$.

\bigskip
\centerline  {\bf 6. Quantum manifolds}
\bigskip

We have discussed how to introduce a dependence of
elements of quantum group from classical parameters.
One can compare this with a map from a classical manifold to
the Grassmann algebra, $x\to \eta (x)$, where
$$
\eta (x) \eta (y) + \eta (y) \eta (x) =0
$$
It is rather natural to consider in this context
also a noncommutative generalization of supermanifolds.
We will make here only a few simple remarks.
We have discussed in the Introduction two points
of view in algebraic geometry: the geometrical
approach and the functorial one. Accordingly  there
are two approaches to supermanifolds. About the geometrical
approach see for example \cite{51,52} and about the
functorial one , for example, \cite{53}-\cite{55}. Therefore
one can do two corresponding approaches in the noncommutative
case which will be, generally, not necessary equivalent
each to other. One can define quantum manifold in any
of these approaches.

Supermanifold $M^{m,n}$, see for example \cite{51,52},
has local coordinates $(x^{\mu},\eta^a)$,
$\mu =1,...m, a=1,...n$, where $x^{\mu}$ are commutative
coordinates, $x^{\mu}x^{\nu}=x^{\nu}x^{\mu}$
and $\eta^a$ are anicommutative ones,
$\eta^a \eta^b=-\eta^b \eta^a$.
A natural generalization of supermanifold is a quantum manifold
with coordinates $(x^{\mu},\eta^a)$ where $x^a$ are again
commutative ones and $\eta^a$are generators of a noncommutative algebra.
In particular one can take $\eta^a$ satisfying the relations
$$
\eta^a \eta^b=R^{ab}_{ij}\eta^i\eta^j  + u^{ab}
$$
where $R^{ab}_{ij}$ and $u ^{ab}$ are numbers.
This case includes CCR, quantum
groups and Clifford algerbas.

Quantum manifold is defined by means of a sheaf
of noncommutative algebra with elements
$$
f(x,\eta)=\sum f_{\alpha} (x)\eta^{\alpha},
$$
where $\alpha $ is a multiindex and the sum could be infinite.
It is a generalization
of supermanifold and it is also generalizes the
Kaluza-Klein approach.
If one considers the simplest case of quantum manifold
$M^{m,2}_q$ with $m$ commutative $x^{\mu},\mu =1,...,m$
coordinates and 2 $q$-commutative ones (Manin's plane)\cite{49,50}:
$$
\eta^1 \eta^2 =q\eta^2 \eta^1
$$
then one sees that the quantum manifold
interpolates between a supermanifold, when
$q=-1$ and the Kaluza-Klein type approach, when
$q=1$ and one gets two extra commuting coordinates.
Note that because quantum manifold has
commuting coordinates it is different from
so called super quantum groups and also
different from the A.Connes noncommutative space.
The physical picture behind quantum manifold
is essentially different one.
Perhaps a proper analogy for quantum
manifold is an adele, with real and p-adic
coordinates, see \cite{56} and  \cite{39} for referenses to
a physical discussion.

\bigskip
\centerline  {\bf 7. Discussion and Conclusion}
\bigskip
We have constructed the quantum sheaf ${\cal A}$  of the Hopf algebras
parametrized  by complex square integrable functions $f$ and real-valued
functions $\theta (x)$.
It would be interesting to characterize the sheaf ${\cal A}$
from the perspective of the general sheaf theory   \cite{34}.
It seems that
the theory of observed algebras \cite{15,16} could be suitable for an
invariant description of the situation.

There are considerations of formal differential calculi on quantum group at
purely algebraic level without referring to the dependence of elements
$g(x)$ of quantum group on classical space-time coordinates $x$
\cite{17}-\cite{26}.
Having constructed such objects as $g(x)$  we can  define
the differential form $\omega =g^{-1}dg$ where $dg =\frac{\partial g(x)}
{\partial x^{\mu}}dx^{\mu}$. One differentiates here $g(x)$ with respect
to the real variable $x$. The form $\omega$ satisfies to the Maurer-Cartan
equation $d\omega +\omega\omega =0$. It would be interesting to find algebraic
relations between matrix elements of $\omega$  and $g$ and to compare them
with relations discussing in algebraic differential calculi
 \cite{18}-\cite{26}. Our construction is algebraically selfconsistent
because $g$ and $dg$ are expressed by means of "preonic" fields
$\alpha (x)$, $\alpha ^{*}(x)$. A natural next step would be in developing
a more general theory of quantum sheaves and quantum bundles.

The simplest action for the quantum group chiral field  \cite{1,3,10}
would  be

$$ 
S=\int dx \mbox{Tr}_q\partial_{\mu}g^{-1}\partial_{\mu}g
$$    

An appropriate formalism how to deal with operator variational calculi
was suggested by Adler \cite{35}. We will not discuss the variational
calculus here. Instead one can postulate some differential
equations.

For example, for the q-deformed WZNW-model \cite{1} one has
$$
\partial_u(g^{-1}\partial_v g)=0.
$$
The solution of these equations is
$$
g=g_1 (u)g_2 (v)
$$
with arbitrary smooth functions $g_1$ and $g_2$ taking values
in quantum group and depending on real parameters $u$ and $v$.

Quantum white noise calculus  \cite{36}-\cite{38}
seems to be relevant for investigation of dynamical
equations in quantum group
field theory. Note that we distinguish between $q$-deformation
and $h$-deformation (quantization), for a discussion see \cite{39}.

One can apply the localization
procedure similar to that has been discussed
in this note also to the bosonization of the q-deformation of universal
enveloping
of the Lie algebra  \cite{13}-\cite{15}. It seems  that this approach
is suitable for construction  quantum group
gauge field  and might be
used for treating quantum group dynamics
  \cite{39}-\cite{43}.
 This subject and also differential calculus along group schemes
 \cite{44}-\cite{46},\cite{57} will be discussed in  a
forthcoming paper.

$$~$$
{\bf ACKNOWLEDGMENT}
$$~$$
The author is grateful to I.Ya.Aref'eva, G.E.Arutyunov,
L.Gerritzen, S.V.Kozyrev, $~$
P.P.Kulish, P.B.Medvedev, R.Parthasarathy,
K.S.Viswanathan and E.I.Zelenov
 for stimulating discussions. This work is supported in part by the
 Russian Fund of Fundamental Research.
$$~$$

\bigskip
\end{document}